# Symmetry and Integrability of a Reduced, 3-Dimensional Self-Dual Gauge Field Model

C. J. Papachristou

*Department of Physical Sciences, Naval Academy of Greece, Piraeus 18539, Greece*

papachristou@snd.edu.gr

**Abstract:** A 3-dimensional reduction of the self-dual Yang-Mills (SDYM) equation, named SDYM3, is examined from the point of view of its symmetry and integrability characteristics. By using a non-auto-Bäcklund transformation, this equation is connected to its potential form (PSDYM3) and a certain isomorphism between the Lie algebras of symmetries of the two systems is shown to exist. This isomorphism allows us to study the infinite-dimensional Lie algebraic structure of the "potential symmetries" of SDYM3 by examining the symmetry structure of PSDYM3 (which is an easier task). By using techniques described in a recent paper, the recursion operators for both SDYM3 and PSDYM3 are derived. Moreover, a Lax pair and an infinite set of nonlocal conservation laws for SDYM3 are found, reflecting the fact that SDYM3 is a totally integrable system. This system may physically represent gravitational fields or chiral fields.

*Keywords: Nonlinear Equations; Symmetries; Conservation Laws; Lax Pair; Recursion Operator; Lie Algebras.*

## 1. Introduction

In a recent paper [1] we proposed a scheme by which symmetry and integrability aspects of a certain class of nonlinear partial differential equations (PDEs) are interrelated. We showed how, by starting with the symmetry condition of a PDE, one may derive integrability characteristics such as a Lax pair and an infinite set of (typically nonlocal) conservation laws. Moreover, we described an algorithm for constructing a recursion operator which, in principle, produces an infinite number of symmetries of the PDE from any given one.

As examples, we applied these ideas to two systems of physical interest: the two-dimensional chiral field equation and the full, 4-dimensional self-dual Yang-Mills (SDYM) equation. The former system is a 2-dimensional reduction of the latter, thus shares some of its properties. However, there are differences: Although the SDYM recursion operator produces infinite sets of nontrivial symmetries when acting on both internal and coordinate symmetry transformations [2, 3], the chiral-field recursion operator yields infinite sets of internal symmetries only.

In this paper, we study an intermediate model which represents a 3-dimensional reduction of SDYM. We will name it SDYM3. With appropriate adjustments, this model may describe physical systems such as the complexified Ernst equation [4] or the 3-dimensional chiral field equation [5]. Happily, some important symmetry properties of SDYM, which are absent in the 2-dimensional chiral-field model, are restored in the 3-dimensional case. Thus, the SDYM3 model possesses infinite sets of nontrivial symmetries on both the base space (coordinate symmetries) and the fiber space (internal symmetries).

The Lie algebraic structure of symmetries of SDYM3 is certainly of interest. Although this aspect of the problem will be treated in full in a subsequent paper, some basic ideas are presented here. In the spirit of a recent paper on SDYM [6], we employ the concept of a Bäcklund transformation (BT) to connect SDYM3 with its counterpart in potential form, to be called PSDYM3. This BT also allows one to connect symmetries and recursion operators of the two systems. In particular, the symmetries of PSDYM3 yield "potential symmetries" [6-8] of SDYM3. It is proven that a Lie algebra isomorphism exists between the symmetries of SDYM3 and those of PSDYM3. Thus, to determine the Lie algebraic structure of symmetries of the former system, it suffices to study the corresponding structure of the latter system. This is not just a matter of academic significance but is important for practical reasons also, given that, as will be seen, the PSDYM3 recursion operator is simpler in form compared to the corresponding SDYM3 operator, with the result that the various commutation relations are easier to handle in the PSDYM3 case.

The paper is organized as follows:

In Section 2, the SDYM3-PSDYM3 system and its symmetry conditions are presented.

In Sec.3, the SDYM3 recursion operator is found in the form of a BT for the linear symmetry condition. This operator produces, in principle, an infinite number of symmetry characteristics, which are also seen to be conserved "charges" for SDYM3. A Lax pair for this PDE is also found.

In Sec.4, it is shown that the symmetries of PSDYM3 can be used to construct potential symmetries for SDYM3.

In Sec.5, a Lie algebra isomorphism is shown to exist between the symmetries of SDYM3 and those of PSDYM3. The practical usefulness of this isomorphism is explained.

The concept of isomorphically related (equivalent) recursion operators [6] is introduced in Sec.6. It is proven that the SDYM3 and PSDYM3 recursion operators are equivalent, thus they produce isomorphic symmetry subalgebras for the respective PDEs.

Finally, in Sec.7 we study the existence of infinite-dimensional abelian subalgebras of symmetries of PSDYM3, thus also of SDYM3. The presence of such algebras is a typical characteristic of integrable systems.

To facilitate the reader, we include an Appendix which contains definitions of the key concepts of the total derivative and the Fréchet derivative. For fuller and more rigorous definitions, the reader is referred to the book by Olver [9].

## 2. The SDYM3 – PSDYM3 System

We write the SDYM3 equation in the form

$$F[J] \equiv D_{\bar{y}}(J^{-1}J_y) + D_z(J^{-1}J_z) = 0 \qquad (1)$$

(the bracket notation is explained in the Appendix). We denote by $x^\mu \equiv y, z, \bar{y}$ ($\mu = 1, 2, 3$) the independent variables (assumed complex) and by $D_y, D_z, D_{\bar{y}}$ the total derivatives with respect to these variables. These derivatives will also be denoted by using subscripts [a mixed notation appears in Eq.(1)]. We assume that $J$ is $SL(N,C)$-valued (i.e., $\det J = 1$).

We consider the non-auto-BT

$$J^{-1}J_y = X_z, \quad J^{-1}J_z = -X_{\bar{y}} \qquad (2)$$

The integrability condition $(X_{\bar{y}})_z = (X_z)_{\bar{y}}$ yields the SDYM3 equation (1). The integrability condition $(J_y)_z = (J_z)_y$, which is equivalent to

$$D_y(J^{-1}J_z) - D_z(J^{-1}J_y) + [J^{-1}J_y, J^{-1}J_z] = 0,$$

yields a nonlinear PDE for the "potential" $X$ of Eq.(1), called the "potential SDYM3 equation" or PSDYM3:

$$G[X] \equiv X_{y\bar{y}} + X_{zz} + [X_z, X_{\bar{y}}] = 0 \qquad (3)$$

Noting that, according to Eq.(2), $(trX)_z = [tr(\ln J)]_y = [\ln(\det J)]_y$, etc., we see that the condition $\det J = 1$ can be satisfied by requiring that $trX=0$ [this requirement is compatible with the PSDYM3 equation (3)]. Hence, $SL(N,C)$ SDYM3 solutions correspond to $sl(N,C)$ PSDYM3 solutions.

At this point we introduce the covariant derivative operators

$$\hat{A}_y \equiv D_y + [J^{-1}J_y, \ ] = D_y + [X_z, \ ]$$
$$\hat{A}_z \equiv D_z + [J^{-1}J_z, \ ] = D_z - [X_{\bar{y}}, \ ]$$

where the BT (2) has been taken into account. By using Eq.(3) and the Jacobi identity, the zero-curvature condition $[\hat{A}_y, \hat{A}_z]=0$ is shown to be satisfied, as expected in view of the fact that the "connections" $J^{-1}J_y$ and $J^{-1}J_z$ are pure gauges. Moreover, the above operators are derivations on the Lie algebra of $sl(N,C)$-valued functions, satisfying a Leibniz rule of the form

$$\hat{A}_y[M,N] = [\hat{A}_yM, N] + [M, \hat{A}_yN]$$
$$\hat{A}_z[M,N] = [\hat{A}_zM, N] + [M, \hat{A}_zN]$$

for any matrix functions $M, N$.

Let $\delta J = \alpha Q[J]$ and $\delta X = \alpha \Phi[X]$ be infinitesimal symmetries of Eqs.(1) and (3), respectively ($\alpha$ is an infinitesimal parameter), with corresponding symmetry characteristics $Q$ and $\Phi$. (We note that *any* symmetry of a PDE can be expressed as a transformation of the dependent variable alone [7, 9], i.e., is equivalent to a "vertical" symmetry.) We will denote by $\Delta M[J]$ the Fréchet derivative (see Appendix) of a function $M$ with respect to the characteristic $Q$. Similarly, by $\Delta N[X]$ we will denote the Fréchet derivative of a function $N$ with respect to $\Phi$. In particular, $\Delta J = Q$ and $\Delta X = \Phi$. The symmetry conditions for the PDEs (1) and (3) are, respectively,

$$\Delta F[J] = 0 \mod F[J] \quad \text{and} \quad \Delta G[X] = 0 \mod G[X] \, .$$

By using the commutativity of the Fréchet derivative with total derivatives (see Appendix), and the fact that

$$\Delta(J^{-1}J_y) = \hat{A}_y(J^{-1}Q), \quad \Delta(J^{-1}J_z) = \hat{A}_z(J^{-1}Q) \, ,$$

the first of the above conditions leads to a linear PDE for the characteristic $Q$:

$$S(Q;J) \equiv (D_{\bar{y}}\hat{A}_y + D_z\hat{A}_z)(J^{-1}Q) = 0 \mod F[J] \tag{4}$$

which represents the symmetry condition for SDYM3.

The symmetry condition for PSDYM3 reads:

$$(\hat{A}_y D_{\bar{y}} + \hat{A}_z D_z)\Phi = 0 \, .$$

By using the operator identity

$$\begin{aligned}\hat{A}_y D_{\bar{y}} + \hat{A}_z D_z &= D_{\bar{y}}\hat{A}_y + D_z\hat{A}_z - [F[J], \ ] \\ &= D_{\bar{y}}\hat{A}_y + D_z\hat{A}_z \mod F[J]\end{aligned} \tag{5}$$

we get the linear PDE for $\Phi$:

$$S(\Phi;X) \equiv (D_{\bar{y}}\hat{A}_y + D_z\hat{A}_z)\Phi = 0 \mod G[X] \tag{6}$$

By comparing Eqs.(4) and (6), we notice that $J^{-1}Q$ and $\Phi$ satisfy the same PDE. Hence, we conclude that

*if $Q$ is an SDYM3 symmetry characteristic, then $\Phi = J^{-1}Q$ is a PSDYM3 characteristic.*

Conversely,

*if $\Phi$ is a PSDYM3 symmetry characteristic, then $Q = J\Phi$ is an SDYM3 characteristic.*

## 3. Recursion Operator, Conserved Charges, and Lax Pair

We seek a recursion operator [9] for SDYM3, i.e., a linear operator which produces new symmetry characteristics $Q'$ from "old" ones, $Q$. As in [1], we want to express this operator in the form of an auto-BT for the linear PDE (4) (which represents the symmetry condition for SDYM3). Moreover, this transformation must be consistent with the physical requirement $tr(J^{-1}Q) = 0$ (i.e., $Q'$ must satisfy this property if $Q$ does), which is necessary in order that the *SL(N,C)* character of the SDYM3 solution be preserved.

The auto-BT for the PDE (4) is similar to that found in [1] for SDYM. Specifically,

$$\hat{A}_y(J^{-1}Q) = (J^{-1}Q')_z , \quad \hat{A}_z(J^{-1}Q) = -(J^{-1}Q')_{\bar{y}} \tag{7}$$

Integrability for $Q'$ requires that $Q$ satisfy Eq.(4). Integrability for $Q$, expressed by the condition $[\hat{A}_y, \hat{A}_z](J^{-1}Q) = 0$, and upon using the operator identity (5), leads us again to Eq.(4), this time for $Q'$. The BT (7) may be regarded as an *invertible recursion operator* for the SDYM3 equation. It can be re-expressed as

$$\begin{aligned}\hat{A}_y\left(J^{-1}Q^{(n)}\right) &= D_z\left(J^{-1}Q^{(n+1)}\right) \\ \hat{A}_z\left(J^{-1}Q^{(n)}\right) &= -D_{\bar{y}}\left(J^{-1}Q^{(n+1)}\right)\end{aligned} \tag{8}$$

($n = 0, \pm 1, \pm 2, ...$). From this we get a doubly infinite set of *nonlocal conservation laws* of the form

$$(D_{\bar{y}}\hat{A}_y + D_z\hat{A}_z)\left(J^{-1}Q^{(n)}\right) = 0 \mod F[J] \tag{9}$$

where the "conserved charges" $Q^{(n)}$ are symmetry characteristics.

Finally, the *Lax pair* for SDYM3, analogous to that found in [1] for SDYM, is

$$D_z(J^{-1}\Psi) = \lambda\hat{A}_y(J^{-1}\Psi) , \quad D_{\bar{y}}(J^{-1}\Psi) = -\lambda\hat{A}_z(J^{-1}\Psi) \tag{10}$$

(where $\lambda$ is a complex "spectral" parameter). The proof of the Lax-pair property is sketched as follows: By the integrability condition $(J^{-1}\Psi)_{z\bar{y}} - (J^{-1}\Psi)_{\bar{y}z} = 0$, we get:

$$S(\Psi;J) \equiv (D_{\bar{y}}\hat{A}_y + D_z\hat{A}_z)(J^{-1}\Psi) = 0 .$$

On the other hand, the integrability condition $[\hat{A}_y, \hat{A}_z](J^{-1}\Psi) = 0$, by using the operator identity (5), yields:

$$S(\Psi;J) - \left[F[J], J^{-1}\Psi\right] = 0 .$$

Therefore, $\left[F[J], J^{-1}\Psi\right] = 0$. This is valid independently of $\Psi$ if $F[J]=0$, i.e., if $J$ is an SDYM3 solution. We conclude that the linear system (10) is a Lax pair for the SDYM3 equation (1), the solution $\Psi$ of which pair is a symmetry characteristic satisfying Eq.(4): $S(\Psi;J) = 0$. This Lax pair is different from that found by Nakamura for the Ernst equation [4].

## 4. Potential Symmetries of SDYM3

We recall that every SDYM3 symmetry characteristic can be expressed as $Q=J\Phi$, where $\Phi$ is a PSDYM3 characteristic. Let $\Phi$ be a characteristic which depends locally or nonlocally on $X$ and/or various derivatives of $X$. By the BT (2), $X$ must be an integral of $J$ and its derivatives, and so it and its derivative $X_y$ are nonlocal in $J$. On the other hand, according to Eq.(2), the quantities $X_{\bar{y}}$ and $X_z$ depend locally on $J$. Thus, in general, $\Phi$ can be local or nonlocal in $J$. In the case where $\Phi$ is nonlocal in $J$, we say that the characteristic $Q=J\Phi$ expresses a *potential symmetry* of SDYM3 [7, 8].

Clearly, to obtain the complete set of potential symmetries of SDYM3, one must first find the totality of symmetries of PSDYM3. To this end, we need the recursion operator for the latter system. Having found the analogous operator for SDYM3, expressed by the BT (7), and by using the fact that $\Phi = J^{-1}Q$ is a PSDYM3 symmetry characteristic when $Q$ is an SDYM3 characteristic, we easily get the recursion operator for PSDYM3 in the form of a BT for the symmetry condition (6):

$$\hat{A}_y \Phi = \Phi'_z \ , \quad \hat{A}_z \Phi = -\Phi'_{\bar{y}} \qquad (11)$$

Since the above two PDEs are consistent with each other, we can use the first one to write $\Phi' = \hat{R}\Phi$, where we have introduced the linear operator

$$\hat{R} = D_z^{-1} \hat{A}_y \qquad (12)$$

To show that the operator (12) is indeed a recursion operator for PSDYM3, we consider a symmetry characteristic $\Phi$ of Eq.(3), i.e., a solution of Eq.(6): $S(\Phi;X)=0$. Then, by using the operator identity (5), and by taking into account the commutativity of covariant derivatives (zero-curvature condition), we have:

$$\begin{aligned}
S(\Phi';X) = S(\hat{R}\Phi;X) &= (\hat{A}_y D_{\bar{y}} + \hat{A}_z D_z) \hat{R}\Phi \\
&= \hat{A}_y D_z^{-1} D_{\bar{y}} \hat{A}_y \Phi + \hat{A}_z \hat{A}_y \Phi \\
&= \hat{A}_y D_z^{-1} (D_{\bar{y}} \hat{A}_y + D_z \hat{A}_z)\Phi + [\hat{A}_z, \hat{A}_y]\Phi \\
&= \hat{A}_y D_z^{-1} S(\Phi;X) + [\hat{A}_z, \hat{A}_y]\Phi = 0
\end{aligned}$$

which proves that $\Phi' = \hat{R}\Phi$ is a symmetry when $\Phi$ is a symmetry.

For $sl(N,C)$ PSDYM3 solutions, the symmetry characteristic $\Phi$ must be traceless. Then, so will be the characteristic $\Phi' = \hat{R}\Phi$. That is, the recursion operator (12) preserves the $sl(N,C)$ character of PSDYM3.

As is easy to see, any power $\hat{R}^n$ ($n=0,\pm 1, \pm 2, \cdots$) of an invertible recursion operator also is a recursion operator. Thus, given any symmetry characteristic $\Phi^{(0)}$, one may obtain, in principle, an infinite set of characteristics:

$$\Phi^{(n)} = \hat{R}\Phi^{(n-1)} = \hat{R}^n \Phi^{(0)} \qquad (n=0,\pm 1, \pm 2, \cdots) \qquad (13)$$

Let us see some examples of using the recursion operator (12) to find PSDYM3 symmetries and corresponding SDYM3 potential symmetries:

1. Take $\Phi^{(0)} = M$, where $M$ is a constant, traceless matrix. Then,

$$\Phi^{(1)} = \hat{R}\Phi^{(0)} = [X, M] \ .$$

The corresponding potential symmetry of SDYM3 is

$$Q^{(1)} = J\Phi^{(1)} = J[X, M] \ .$$

(This is nonlocal in $J$ due to the presence of $X$.) Higher-order potential symmetries are found recursively by repeated application of the recursion operator (12). We thus obtain an infinite sequence of "internal" symmetries (i.e., symmetry transformations in the "fiber" space), of the form:

$$Q^{(n)} = J\Phi^{(n)} = J\hat{R}^n\Phi^{(0)} \quad (n = 0, 1, 2, \cdots) \tag{14}$$

In the case of the complexified Ernst equation, these are precisely the internal symmetries found by Nakamura [4].

2. Take $\Phi^{(0)} = X_y$, which represents a coordinate symmetry (symmetry transformation in the "base" space of the independent variables $x^\mu$), specifically, invariance under $y$-translation. By applying the recursion operator, we get:

$$\Phi^{(1)} = \hat{R}\Phi^{(0)} = D_z^{-1}(X_{yy} + [X_z, X_y]) = D_z^{-1}(X_{yy} + [J^{-1}J_y, X_y]) \ .$$

Both $\Phi^{(0)}$ and $\Phi^{(1)}$ are nonlocal in $J$ (due to the presence of the $y$-derivatives of $X$, as well as of the integral operator with respect to $z$). We thus obtain the potential symmetries of SDYM3,

$$Q^{(0)} = J\Phi^{(0)} = JX_y \ ,$$
$$Q^{(1)} = J\Phi^{(1)} = JD_z^{-1}(X_{yy} + [J^{-1}J_y, X_y]) \ .$$

We note that, by applying the recursion operator to the translational characteristics $\Phi^{(0)} = X_z$ and $\Phi^{(0)} = X_{\bar{y}}$ [both of which are *local* in $J$, in view of the BT (2)], we get, respectively, $\Phi^{(1)} = X_y$ (which is nothing new) and $\Phi^{(1)} = -X_z$ (again, nothing new).

3. Take $\Phi^{(0)} = yX_y + zX_z + \bar{y}X_{\bar{y}}$, which represents a scale change of the $x^\mu$. This is nonlocal in $J$ due to $X_y$. We leave it to the reader to show that

$$\Phi^{(1)} = \hat{R}\Phi^{(0)} = zX_y - \bar{y}J^{-1}J_y + yD_z^{-1}(X_{yy} + [J^{-1}J_y, X_y]) \ ,$$

where the PSDYM3 equation (3) and the BT (2) have been taken into account. This is also nonlocal in $J$. We conclude that $J\Phi^{(0)}$ and $J\Phi^{(1)}$ are potential symmetries for SDYM3.

## 5. Lie Algebra Isomorphism

We now study the connection between the Lie algebras of symmetries of SDYM3 and PSDYM3. If these algebras are isomorphic, then any Lie algebraic conclusion regarding the PSDYM3 equation will also be true for the SDYM3 equation. What is the practical value of this? As we saw, the recursion operator for PSDYM3 is given by Eq.(12):

$$\hat{R} = D_z^{-1} \hat{A}_y \; .$$

On the other hand, from the BT (7) we get, by using the first equation,

$$Q' = J D_z^{-1} \hat{A}_y (J^{-1} Q) = J \hat{R} (J^{-1} Q) \equiv \hat{T} Q$$

where $\hat{T}$ is the operator form of the SDYM3 recursion operator:

$$\hat{T} = J D_z^{-1} \hat{A}_y J^{-1} = J \hat{R} J^{-1} \qquad (15)$$

Obviously, $\hat{R}$ is of a simpler form compared to $\hat{T}$. Accordingly, the Lie algebraic structure of the infinite sequences of symmetries generated by the former operator will be easier to study compared to the corresponding structure of symmetries produced by the latter operator. So, we are seeking an isomorphism between the Lie algebras of symmetries of the PDEs (1) and (3).

In the spirit of [6], where the 4-dimensional SDYM case was studied, we consider the pair of PDEs:

$$\hat{A}_y (J^{-1} Q) = \Phi_z \; , \qquad \hat{A}_z (J^{-1} Q) = -\Phi_{\bar{y}} \qquad (16)$$

Equation (16) is a BT connecting the symmetry characteristic $\Phi$ of PSDYM3 with the symmetry characteristic $Q$ of SDYM3. Indeed, the integrability condition $(\Phi_z)_{\bar{y}} = (\Phi_{\bar{y}})_z$ yields the symmetry condition (4) for SDYM3, while the integrability condition $[\hat{A}_y, \hat{A}_z](J^{-1} Q) = 0$, valid in view of the zero-curvature condition, yields the PSDYM3 symmetry condition (6). Please note carefully that the $Q$ and $\Phi$ in Eq.(16) are *not* related by the simple algebraic relation $Q = J\Phi$. Note also that the system (16) is compatible with the constraints that $\Phi$ and $J^{-1}Q$ be traceless, as required for producing $sl(N,C)$ PSDYM3 solutions and $SL(N,C)$ SDYM3 solutions, respectively.

We observe that, for a given $Q$, the solution of the BT (16) for $\Phi$ is not unique, and neither is the solution for $Q$, for a given $\Phi$. Indeed, in either case the solution may contain arbitrary additive terms. We normalize the process by agreeing to ignore such terms, so that, in particular, the characteristic $Q=0$ corresponds to the characteristic $\Phi=0$. In this way, the BT (16) establishes a one-to-one correspondence between the symmetries of SDYM3 and those of PSDYM3. We will now show that this correspondence is a Lie algebra isomorphism.

**Lemma:** The Fréchet derivative $\Delta$ with respect to the characteristic $\Phi$, and the recursion operator $\hat{R}$ of Eq.(12), satisfy the commutation relation

$$[\Delta, \hat{R}] = D_z^{-1}[\Phi_z, \ ] \tag{17}$$

where $\Phi = \Delta X$.

**Proof:** Introducing an auxiliary matrix function $M$, and using the derivation property of $\Delta$ and the commutativity of $\Delta$ with all total derivatives (as well as all powers of such derivatives), we have:

$$\Delta \hat{R} M = \Delta D_z^{-1} \hat{A}_y M = D_z^{-1} \Delta (D_y M + [X_z, M])$$
$$= D_z^{-1}(D_y \Delta M + [(\Delta X)_z, M] + [X_z, \Delta M])$$
$$= D_z^{-1}(\hat{A}_y \Delta M + [\Phi_z, M]) = \hat{R} \Delta M + D_z^{-1}[\Phi_z, M],$$

from which there follows (17).

Now, by the first equation of the BT (16), we can write:

$$\Phi = D_z^{-1} \hat{A}_y (J^{-1} Q) = \hat{R}(J^{-1} Q) \tag{18}$$

Equation (18) defines a linear map from the set of symmetries $Q = \Delta J$ of the PDE (1) to the set of symmetries $\Phi = \Delta X$ of the PDE (3). With the normalization conventions mentioned earlier, this map can be considered invertible, thus constituting a one-to-one correspondence between the symmetries of SDYM3 and those of PSDYM3, for any given solutions $J$ and $X$ connected to each other by the BT (2). Calling this map $I$, we write:

$$I: \Phi = I\{Q\} = \hat{R} J^{-1} Q \quad or \quad \Delta X = I\{\Delta J\} = \hat{R} J^{-1} \Delta J \tag{19}$$

[*Note:* We may omit parentheses, such as those in Eq.(18), by agreeing that an operator acts on the entire expression (e.g., product of functions) on its right, not just on the function adjacent to it. Hence, $\hat{P} M N \equiv \hat{P}(MN)$.]

**Proposition 1:** The map $I$ defined by Eq.(19) is an isomorphism between the symmetry Lie algebras of SDYM3 and PSDYM3.

**Proof:** Consider a pair of symmetries of Eq.(1), indexed by $i$ and $j$, generated by the characteristics $Q^{(l)} = \Delta^{(l)} J$, where $l = i, j$. Similarly, consider a pair of symmetries of Eq.(3), generated by $\Phi^{(l)} = \Delta^{(l)} X$ ($l = i, j$). Further, assume that

$$\Phi^{(l)} = I\{Q^{(l)}\} \quad or \quad \Delta^{(l)} X = I\{\Delta^{(l)} J\}\ .$$

That is,

$$\Phi^{(l)} = \Delta^{(l)} X = \hat{R} J^{-1} \Delta^{(l)} J = \hat{R} J^{-1} Q^{(l)} \; ; \quad l = i, j \tag{20}$$

By the Lie-algebraic property of symmetries of PDEs, the functions $[\Delta^{(i)}, \Delta^{(j)}] J$ and $[\Delta^{(i)}, \Delta^{(j)}] X$ also are symmetry characteristics for SDYM3 and PSDYM3, respectively, where we have put

$$[\Delta^{(i)}, \Delta^{(j)}] J \equiv \Delta^{(i)} \Delta^{(j)} J - \Delta^{(j)} \Delta^{(i)} J = \Delta^{(i)} Q^{(j)} - \Delta^{(j)} Q^{(i)},$$
$$[\Delta^{(i)}, \Delta^{(j)}] X \equiv \Delta^{(i)} \Delta^{(j)} X - \Delta^{(j)} \Delta^{(i)} X = \Delta^{(i)} \Phi^{(j)} - \Delta^{(j)} \Phi^{(i)}.$$

We must now show that

$$[\Delta^{(i)}, \Delta^{(j)}] X = I \{ [\Delta^{(i)}, \Delta^{(j)}] J \} = \hat{R} J^{-1} [\Delta^{(i)}, \Delta^{(j)}] J \tag{21}$$

Putting $l=j$ into Eq.(20), and applying the Fréchet derivative $\Delta^{(i)}$, we have:

$$\Delta^{(i)} \Delta^{(j)} X = \Delta^{(i)} \hat{R} J^{-1} Q^{(j)} = [\Delta^{(i)}, \hat{R}] J^{-1} Q^{(j)} + \hat{R} \Delta^{(i)} J^{-1} Q^{(j)}$$
$$= D_z^{-1} [\Phi_z^{(i)}, J^{-1} Q^{(j)}] + \hat{R} \Delta^{(i)} J^{-1} Q^{(j)},$$

where we have used the commutation relation (17). By Eq.(20),

$$\Phi_z^{(i)} = D_z \hat{R} J^{-1} Q^{(i)} = \hat{A}_y J^{-1} Q^{(i)}.$$

Moreover, by the properties of the Fréchet derivative listed in the Appendix,

$$\Delta^{(i)} J^{-1} Q^{(j)} = -J^{-1} (\Delta^{(i)} J) J^{-1} Q^{(j)} + J^{-1} \Delta^{(i)} Q^{(j)}$$
$$= -J^{-1} Q^{(i)} J^{-1} Q^{(j)} + J^{-1} \Delta^{(i)} Q^{(j)}.$$

So,

$$\Delta^{(i)} \Delta^{(j)} X = D_z^{-1} [\hat{A}_y J^{-1} Q^{(i)}, J^{-1} Q^{(j)}] - \hat{R} J^{-1} Q^{(i)} J^{-1} Q^{(j)} + \hat{R} J^{-1} \Delta^{(i)} Q^{(j)}.$$

Subtracting from this the analogous expression for $\Delta^{(j)} \Delta^{(i)} X$, we have:

$$[\Delta^{(i)}, \Delta^{(j)}] X \equiv \Delta^{(i)} \Delta^{(j)} X - \Delta^{(j)} \Delta^{(i)} X$$
$$= D_z^{-1} \left( [\hat{A}_y J^{-1} Q^{(i)}, J^{-1} Q^{(j)}] + [J^{-1} Q^{(i)}, \hat{A}_y J^{-1} Q^{(j)}] \right)$$
$$- \hat{R} [J^{-1} Q^{(i)}, J^{-1} Q^{(j)}] + \hat{R} J^{-1} (\Delta^{(i)} Q^{(j)} - \Delta^{(j)} Q^{(i)})$$
$$= D_z^{-1} \hat{A}_y [J^{-1} Q^{(i)}, J^{-1} Q^{(j)}] - \hat{R} [J^{-1} Q^{(i)}, J^{-1} Q^{(j)}]$$
$$+ \hat{R} J^{-1} (\Delta^{(i)} \Delta^{(j)} J - \Delta^{(j)} \Delta^{(i)} J)$$
$$= \hat{R} J^{-1} [\Delta^{(i)}, \Delta^{(j)}] J$$

where we have used the derivation property of $\hat{A}_y$. Thus, Eq.(21) has been proven.

## 6. Isomorphically Related Recursion Operators

Following [6], we now introduce the concept of *isomorphically related* (*I-related*) *recursion operators*. Let $\hat{S}$ be a recursion operator for the SDYM3 equation (1) [not necessarily that of Eq.(15)], and let $\hat{P}$ be a recursion operator for the PSDYM3 equation (3) [not necessarily that of Eq.(12)].

**Definition:** The linear operators $\hat{P}$ and $\hat{S}$ will be called *equivalent with respect to the isomorphism I* (or *I-equivalent*, or *I-related*) if the following condition is satisfied:

$$\hat{P}\Phi = I\{\hat{S}Q\} \quad \text{when} \quad \Phi = I\{Q\} \tag{22}$$

where $Q$ and $\Phi$ are symmetry characteristics for the PDEs (1) and (3), respectively.

**Proposition 2:** Any *I*-related recursion operators $\hat{P}$ and $\hat{S}$ satisfy the following operator equation on the infinite-dimensional linear space of all SDYM3 symmetry characteristics:

$$\hat{P}\hat{R}J^{-1} = \hat{R}J^{-1}\hat{S} \tag{23}$$

where $\hat{R}$ is the operator defined in Eq.(12).

**Proof:** By Eqs.(19) and (22),

$$\hat{P}\Phi = \hat{R}J^{-1}\hat{S}Q \quad \text{when} \quad \Phi = \hat{R}J^{-1}Q \;\Rightarrow\; \hat{P}\hat{R}J^{-1}Q = \hat{R}J^{-1}\hat{S}Q$$

for all SDYM3 characteristics $Q$.

**Proposition 3:** The recursion operators $\hat{R}$ and $\hat{T}$, defined by Eqs.(12) and (15), are *I*-equivalent.

**Proof:** Simply note that the operator equation (23) is satisfied by putting $\hat{P} = \hat{R}$ and $\hat{S} = \hat{T}$, and by taking Eq.(15) into account.

Now, let $Q^{(0)}$ be some SDYM3 symmetry characteristic, and let $\Phi^{(0)}$ be the *I*-related PSDYM3 characteristic:

$$\Phi^{(0)} = I\{Q^{(0)}\} = \hat{R}J^{-1}Q^{(0)} \tag{24}$$

Consider also the infinite sets of symmetries of the PDEs (1) and (3), respectively:

$$Q^{(n)} = \hat{T}^n Q^{(0)} \;;\; n = 0,1,2,\cdots \tag{25}$$

$$\Phi^{(n)} = \hat{R}^n \Phi^{(0)} \;;\; n = 0,1,2,\cdots \tag{26}$$

**Proposition 4:** If the set (25) generates an infinite-dimensional Lie subalgebra of SDYM3 symmetries, then the set (26) generates an infinite-dimensional Lie subalgebra of PSDYM3 symmetries, isomorphic to the SDYM3 symmetry subalgebra.

**Proof:** Since the operators $\hat{R}$ and $\hat{T}$ are *I*-equivalent, by Eq.(24) we have:

$$\hat{R}\Phi^{(0)} = I\{\hat{T}Q^{(0)}\} \; .$$

By iterating,

$$\hat{R}^n \Phi^{(0)} = I\{\hat{T}^n Q^{(0)}\} \quad or \quad \Phi^{(n)} = I\{Q^{(n)}\} \; ; \; n = 0,1,2,\cdots \qquad (27)$$

Call *V* and *W* the infinite-dimensional linear spaces spanned by the basis functions (25) and (26), respectively. The elements of *V* and *W* are, correspondingly, symmetry characteristics of SDYM3 and PSDYM3. Equation (27) defines an isomorphism between *V* and *W*. By assumption, the characteristics belonging to *V* generate a Lie subalgebra of the complete Lie algebra of symmetries of SDYM3. We must show that the elements of *W* generate an isomorphic subalgebra of PSDYM3 symmetries. To this end, consider two basis elements $Q^{(i)} = \Delta^{(i)}J$ and $Q^{(j)} = \Delta^{(j)}J$ of *V*. From these we construct the Lie bracket,

$$[\Delta^{(i)}, \Delta^{(j)}]J \equiv \Delta^{(i)}\Delta^{(j)}J - \Delta^{(j)}\Delta^{(i)}J = \Delta^{(i)}Q^{(j)} - \Delta^{(j)}Q^{(i)}$$

which is an SDYM3 symmetry characteristic. This characteristic belongs to the subspace *V* (since this space generates a Lie algebra). Now, let

$$\Phi^{(l)} = \Delta^{(l)}X = I\{Q^{(l)}\} \; ; \; l = i,j$$

be the basis elements of *W* which are *I*-related to the $Q^{(l)}$ $(l = i, j)$, in the way dictated by Eq.(27). By Eq.(21),

$$[\Delta^{(i)}, \Delta^{(j)}]X \equiv \Delta^{(i)}\Phi^{(j)} - \Delta^{(j)}\Phi^{(i)} = I\{[\Delta^{(i)}, \Delta^{(j)}]J\} \; .$$

The quantity on the left is a PSDYM3 symmetry characteristic. Given that *I* is a map from *V* to *W*, this characteristic belongs to the subspace *W*. Thus, *W* is closed under the Lie bracket operation, which means that its elements generate a Lie subalgebra of PSDYM3 symmetries. This subalgebra is *I*-related, i.e. isomorphic, to the corresponding subalgebra of SDYM3 symmetries generated by the elements of *V*.

## 7. Infinite-Dimensional Abelian Subalgebras

The study of the complete symmetry Lie algebra of SDYM3 will be the subject of a future paper. Here, we confine ourselves to the existence of infinite-dimensional abelian subalgebras, the presence of which is a typical characteristic of integrable systems. Since the PDEs (1) and (3) constitute a 3-dimensional reduction of the 4-dimensional SDYM-PSDYM system, certain symmetry aspects of the latter system are expected to be present in the former one also. In particular, the PSDYM equation has been shown to possess Kac-Moody symmetry algebras associated with both internal and coordinate transformations [2]. These algebras possess infinite-dimensional abelian subalgebras. Such abelian structures exist for the reduced 3-dimensional system also. The following theorem follows directly from a more general one concerning the 4-dimensional PSDYM equation [2]:

**Theorem:** Consider a PSDYM3 symmetry, having a characteristic of the form

$$\Phi^{(0)} = \Delta^{(0)} X = \hat{L} X$$

where $\hat{L}$ is a linear operator. By repeated application of the recursion operator (12), we construct an infinite sequence of PSDYM3 characteristics,

$$\Phi^{(n)} = \Delta^{(n)} X = \hat{R}^n \Phi^{(0)} = \hat{R}^n \hat{L} X \; ; \; n = 0, 1, 2, \cdots \qquad (28)$$

We assume that the operator $\hat{L}$ obeys the commutation relations

$$[\Delta^{(n)}, \hat{L}] = 0 \quad \text{and} \quad [\hat{L}, \hat{R}] = D_z^{-1} [D_z \hat{L} X, \; ] \; .$$

Then, the set (28) represents an infinite-dimensional abelian symmetry algebra:

$$[\Delta^{(m)}, \Delta^{(n)}] X \equiv \Delta^{(m)} \Phi^{(n)} - \Delta^{(n)} \Phi^{(m)} = 0 \; .$$

We note that the commutation relation (17) is written, in this case,

$$[\Delta^{(n)}, \hat{R}] = D_z^{-1} [D_z \Delta^{(n)} X, \; ] = D_z^{-1} [D_z \hat{R}^n \hat{L} X, \; ] \; .$$

As an example, it can be checked that the conditions of this theorem are satisfied for the linear operators $\hat{L}_1 = D_y$ and $\hat{L}_2 = y D_y + z D_z + \bar{y} D_{\bar{y}}$, corresponding to the PSDYM3 symmetries $\Phi^{(0)} = X_y$ and $\Phi^{(0)} = y X_y + z X_z + \bar{y} X_{\bar{y}}$, respectively. The $I$-related SDYM3 symmetries are $Q^{(0)} = J_y$ and $Q^{(0)} = y J_y + z J_z + \bar{y} J_{\bar{y}}$. We thus obtain two infinite-dimensional abelian subsymmetries of PSDYM3:

$$\Phi^{(n)} = \Delta^{(n)} X = \hat{R}^n X_y \; ; \; n = 0, 1, 2, \cdots$$
$$\Phi^{(n)} = \Delta^{(n)} X = \hat{R}^n (y X_y + z X_z + \bar{y} X_{\bar{y}}) \; ; \; n = 0, 1, 2, \cdots$$

and two *I*-related abelian (by Proposition 4) sybsymmetries of SDYM3:

$$Q^{(n)} = \Delta^{(n)} J = \hat{T}^n J_y \; ; \; n = 0, 1, 2, \cdots$$
$$Q^{(n)} = \Delta^{(n)} J = \hat{T}^n (y J_y + z J_z + \bar{y} J_{\bar{y}}) \; ; \; n = 0, 1, 2, \cdots$$

## 8. Summary

We have explored the symmetry and integrability characteristics of a 3-dimensional reduction of the full 4-dimensional self-dual Yang-Mills system. The former model is physically interesting since, with appropriate adjustments, it may describe chiral fields [5, 10] or axially-symmetric gravitational fields [4]. We have used the techniques described in [1] to derive a recursion operator, a Lax pair, and an infinite set of conserved "charges". We have studied the existence of potential symmetries, and we have investigated certain aspects of the Lie algebraic structure of symmetries of our model. The study of the full symmetry algebra of this model will be the subject of a future paper.

## 9. Appendix

To make this article as self-contained as possible, we define two key concepts that are being used, namely, the total derivative and the Fréchet derivative. The reader is referred to the extensive review article [11] by this author for more details. (It should be noted, however, that our present definition of the Fréchet derivative corresponds to the definition of the *Lie* derivative in that article. Since these two derivatives are *locally* indistinguishable, this discrepancy in terminology should not cause any concern mathematically.)

We consider the set of all PDEs of the form $F[u]=0$, where, for simplicity, the solutions $u$ (which may be matrix-valued) are assumed to be functions of only two variables, $x$ and $t$: $u=u(x,t)$. In general,

$$F[u] \equiv F(x, t, u, u_x, u_t, u_{xx}, u_{tt}, u_{xt}, \cdots) \ .$$

Geometrically, we say that the function $F$ is defined in a *jet space* [9, 12] with coordinates $x$, $t$, $u$, and as many partial derivatives of $u$ as needed for the given problem. A solution of the PDE $F[u]=0$ is then a surface in this jet space.

Let $F[u]$ be a given function in the jet space. When differentiating such a function with respect to $x$ or $t$, both implicit (through $u$) and explicit dependence of $F$ on these variables must be taken into account. If $u$ is a scalar quantity, we define the *total derivative operators* $D_x$ and $D_t$ as follows:

$$D_x = \frac{\partial}{\partial x} + u_x \frac{\partial}{\partial u} + u_{xx} \frac{\partial}{\partial u_x} + u_{xt} \frac{\partial}{\partial u_t} + \cdots$$

$$D_t = \frac{\partial}{\partial t} + u_t \frac{\partial}{\partial u} + u_{xt} \frac{\partial}{\partial u_x} + u_{tt} \frac{\partial}{\partial u_t} + \cdots$$

(note that the operators $\partial/\partial x$ and $\partial/\partial t$ concern only the explicit dependence of $F$ on $x$ and $t$). If, however, $u$ is matrix-valued, the above representation has only symbolic significance and cannot be used for actual calculations. We must therefore define the total derivatives $D_x$ and $D_t$ in more general terms.

We define a linear operator $D_x$, acting on functions $F[u]$ in the jet space and having the following properties:

1. On functions $f(x,t)$ in the base space,

$$D_x f(x,t) = \partial f/\partial x \equiv \partial_x f .$$

2. On functions $F[u] = u$ or $u_x$, $u_t$, etc., in the "fiber" space,

$$D_x u = u_x , \quad D_x u_x = u_{xx} , \quad D_x u_t = u_{tx} = u_{xt} , \text{ etc.}$$

3. The operator $D_x$ is a derivation on the algebra of all functions $F[u]$ in the jet space (i.e., the Leibniz rule is satisfied):

$$D_x (F[u] G[u]) = (D_x F[u]) G[u] + F[u] D_x G[u] .$$

We similarly define the operator $D_t$. Extension to higher-order total derivatives is obvious (although these latter derivatives are no longer derivations, i.e., they do not satisfy the Leibniz rule). The following notation has been used in this article:

$$D_x F[u] \equiv F_x[u] , \quad D_t F[u] \equiv F_t[u] .$$

Finally, it can be shown that, for any matrix-valued functions $A$ and $B$ in the jet space, we have

$$(A^{-1})_x = -A^{-1} A_x A^{-1} , \quad (A^{-1})_t = -A^{-1} A_t A^{-1}$$

and

$$D_x [A,B] = [A_x, B] + [A, B_x] , \quad D_t [A,B] = [A_t, B] + [A, B_t]$$

where square brackets denote commutators.

Let now $\delta u \simeq \alpha Q[u]$ be an infinitesimal symmetry transformation (with characteristic $Q[u]$) for the PDE $F[u]=0$. We define the *Fréchet derivative* with respect to the characteristic $Q$ as a linear operator $\Delta$ acting on functions $F[u]$ in the jet space and having the following properties:

1. On functions $f(x,t)$ in the base space,

$$\Delta f(x,t) = 0$$

(this is a consequence of our liberty to choose all our symmetries to be in "vertical" form [7, 9]).

2. On $F[u] = u$,

$$\Delta u = Q[u] .$$

3. The operator $\Delta$ commutes with total derivative operators of any order.

4. The Leibniz rule is satisfied:

$$\Delta(F[u]G[u]) = (\Delta F[u])G[u] + F[u]\Delta G[u] .$$

The following properties can be proven:

$$\Delta u_x = (\Delta u)_x = Q_x[u] , \quad \Delta u_t = (\Delta u)_t = Q_t[u]$$

$$\Delta(A^{-1}) = -A^{-1}(\Delta A)A^{-1} ; \quad \Delta[A,B] = [\Delta A, B] + [A, \Delta B]$$

where $A$ and $B$ are any matrix-valued functions in the jet space.

If the solution $u$ of the PDE is a scalar function (thus so is the characteristic $Q$), the Fréchet derivative with respect to $Q$ admits a differential-operator representation of the form

$$\Delta = Q\frac{\partial}{\partial u} + Q_x\frac{\partial}{\partial u_x} + Q_t\frac{\partial}{\partial u_t} + Q_{xx}\frac{\partial}{\partial u_{xx}} + Q_{tt}\frac{\partial}{\partial u_{tt}} + Q_{xt}\frac{\partial}{\partial u_{xt}} + \cdots$$

Such representations, however, are not valid for PDEs in matrix form. In these cases we must resort to the general definition of the Fréchet derivative given above.

Finally, by using the Fréchet derivative, the symmetry condition for a PDE $F[u] = 0$ can be expressed as follows [7, 9]:

$$\Delta F[u] = 0 \mod F[u] .$$

This condition yields a linear PDE for the symmetry characteristic $Q$, of the form

$$S(Q;u) = 0 \mod F[u] .$$